\documentclass[]{nature}
\makeatletter\if@twocolumn\PassOptionsToPackage{switch}{lineno}\else\fi\makeatother

\usepackage{tabulary,graphicx,amsmath,amsfonts,amssymb}
\usepackage[utf8x]{inputenc}
\usepackage[T1]{fontenc}

\makeatletter

\renewenvironment{figure*}
               {\@dblfloat{figure}}
               {\end@dblfloat}

\renewenvironment{table*}
               {\@dblfloat{table}}
               {\end@dblfloat}

\makeatother

\usepackage{url,multirow,morefloats,floatflt,cancel,tfrupee}
\makeatletter

\AtBeginDocument{\@ifpackageloaded{textcomp}{}{\usepackage{textcomp}}}
\makeatother
\usepackage{colortbl}
\usepackage{xcolor}
\usepackage{pifont}
\usepackage[nointegrals]{wasysym}
\makeatletter

\def\mcWidth#1{\csname TY@F#1\endcsname+\tabcolsep}

\def\cAlignHack{\rightskip\@flushglue\leftskip\@flushglue\parindent\z@\parfillskip\z@skip}
\def\rAlignHack{\rightskip\z@skip\leftskip\@flushglue \parindent\z@\parfillskip\z@skip}

\@ifundefined{etal}{}{}

\usepackage{ifxetex}
\ifxetex\else\if@twocolumn\@ifpackageloaded{stfloats}{}{\usepackage{dblfloatfix}}\fi\fi

\AtBeginDocument{
\expandafter\ifx\csname eqalign\endcsname\relax
\def\eqalign#1{\null\vcenter{\def\\{\cr}\openup\jot\m@th
  \ialign{\strut$\displaystyle{##}$\hfil&$\displaystyle{{}##}$\hfil
      \crcr#1\crcr}}\,}
\fi
}

\AtBeginDocument{%
  \@ifpackageloaded{endfloat}%
   {\renewcommand\efloat@iwrite[1]{\immediate\expandafter\protected@write\csname efloat@post#1\endcsname{}}}{\newif\ifefloat@tables}%
}%

\def\BreakURLText#1{\@tfor\brk@tempa:=#1\do{\brk@tempa\hskip0pt}}
\let\lt=<
\let\gt=>
\def\processVert{\ifmmode|\else\textbar\fi}

\@ifundefined{subparagraph}{
\def\subparagraph{\@startsection{paragraph}{5}{2\parindent}{0ex plus 0.1ex minus 0.1ex}%
{0ex}{\normalfont\small\itshape}}%
}{}

\newcommand\role[1]{\unskip}
\newcommand\aucollab[1]{\unskip}
  
\@ifundefined{tsGraphicsScaleX}{\gdef\tsGraphicsScaleX{1}}{}
\@ifundefined{tsGraphicsScaleY}{\gdef\tsGraphicsScaleY{.9}}{}
\def\checkGraphicsWidth{\ifdim\Gin@nat@width>\linewidth
	\tsGraphicsScaleX\linewidth\else\Gin@nat@width\fi}

\def\checkGraphicsHeight{\ifdim\Gin@nat@height>.9\textheight
	\tsGraphicsScaleY\textheight\else\Gin@nat@height\fi}

\def\fixFloatSize#1{}
\let\ts@includegraphics\includegraphics

\def\inlinegraphic[#1]#2{{\edef\@tempa{#1}\edef\baseline@shift{\ifx\@tempa\@empty0\else#1\fi}\edef\tempZ{\the\numexpr(\numexpr(\baseline@shift*\f@size/100))}\protect\raisebox{\tempZ pt}{\ts@includegraphics{#2}}}}

\AtBeginDocument{\def\includegraphics{\@ifnextchar[{\ts@includegraphics}{\ts@includegraphics[width=\checkGraphicsWidth,height=\checkGraphicsHeight,keepaspectratio]}}}

\DeclareMathAlphabet{\mathpzc}{OT1}{pzc}{m}{it}

\def\URL#1#2{\@ifundefined{href}{#2}{\href{#1}{#2}}}

\def\UrlOrds{\do\*\do\-\do\~\do\'\do\"\do\-}%
\g@addto@macro{\UrlBreaks}{\UrlOrds}

\edef\fntEncoding{\f@encoding}

\makeatother

\newif\ifmultipleabstract\multipleabstractfalse%
\usepackage[nomarkers,tablesfirst,nolists]{endfloat}

\def\fixFloatSize#1{}
\usepackage{float}

\newcommand{\texttildeapprox}{{\fontfamily{pcr}\selectfont\texttildelow}}

\begin{document}

\title{A Machine-Learning Approach for Earthquake Magnitude Estimation}

\author{S. Mostafa Mousavi$^{1}$\thanks{Corresponding author.}\hspace{.4pc}\thanks{E-mail: mmousavi@stanford.edu}~ \&
              Gregory C. Beroza
    }

\maketitle 

\begin{affiliations}
  \item 
    Geophysics Department\unskip, 
    Stanford University \unskip, Stanford\unskip, California\unskip, USA
\end{affiliations}

\begin{abstract}
In this study we develop a single-station deep-learning approach for fast and reliable estimation of earthquake magnitude directly from raw waveforms. We design a regressor composed of convolutional and recurrent neural networks that is not sensitive to the data normalization, hence waveform amplitude information can be utilized during the training. Our network can predict earthquake magnitudes with an average error close to zero and standard deviation of {\texttildeapprox}0.2 based on single-station waveforms without instrument response correction. We test the network for both local and duration magnitude scales and show a station-based learning can be an effective approach for improving the performance. The proposed approach has a variety of potential applications from routine earthquake monitoring to early warning systems. 

\textbf{ Plain Language Summary}

The size of an earthquake at its source is measured from the amplitude (or sometimes the duration) of the ground motion recorded on seismic instruments, and is expressed in terms of magnitude. Magnitude is a logarithmic measure and usually is measured based on data recorded by multiple stations after applying some pre-proccessing and corrections to the raw signals. Here, we introduce the first successful deep-learning approach to  estimate directly the magnitude from raw seismic signals recorded on a single station. 
\end{abstract}\def\keywordstitle{Keywords}


\section{Key Points:}
A deep-learning approach is presented for earthquake magnitude estimation. 

Network consists of both convolutional and recurrent neural networks.

It can estimate both $M_L $ and $M_d $ directly from raw seismograms recorded on a single station.

\section{Introduction}
Earthquake magnitude is one of the fundamental parameters for earthquake characterization. It is a logarithmic measure that represents the strength of the earthquake source. Magnitude provides the public with quick information on earthquakes and is used in scientific research as well. Since Charles F. Richter introduced the earthquake magnitude scale, the so-called local (\textit{M\ensuremath{_{L}}}) or Richter scale, in 1935 (\cite{598244:13913693}), there have been many studies proposing various types of magnitude scales. These magnitude scales measure different properties of the seismic waves (e.g., low-frequency energy vs. high-frequency energy, surface waves vs. body waves) and although they may represent fundamentally different characteristics of the source, they are suitable for different earthquake sizes and different epicentral distance ranges. Most magnitude scales are empirical. Usually a magnitude \textit{M} is determined from the amplitude \textit{A} and the period \textit{T} of a certain type of seismic wave through a formula that contains several constants. These constants are determined in such a way that the magnitudes for a new scale agree with those of an existing one at least over a certain magnitude range. In some cases, the duration of shaking seismogram is used to determine magnitude\unskip~\cite{598244:13913713}. Hence, the various magnitude-types may have values that differ by more than a magnitude-unit for very large and very small earthquakes as well as for some specific classes of seismic source. This is because the physical process underlying an earthquake is complex\unskip~\cite{598244:13913714}. 

A typical magnitude estimation procedure includes: 1) converting the raw seismograms into displacement after correcting them for the instrument response. 2) estimating magnitudes at each station after correcting for propagation effect, for the station-epicenter distance. 3) averaging the single-station estimates to compensate for possible site effects. In earthquake and tsunami early warning systems (e.g.\unskip~\cite{598244:14121469}), however, where time is of the essence to broadcast a warning, rapid and reliable estimation of a preliminary magnitude with what data is immediatly available has specific importance \unskip~\cite{598244:13913714}. In this study, we present a fast and reliable method for end-to-end estimation of earthquake magnitude from raw seismograms observed at single stations. Although there have been attempts to estimate earthquake magnitude using deep neural networks (e.g. \unskip~\cite{598244:13914097}), to the best of our knowledge, this is the first successful study to do so. 
    
\section{Method}
Neural networks have been shown to be a powerful tool for earthquake signal processing and characterization (e.g. \unskip~\cite{598244:13914619,598244:13914705,598244:13914706,598244:13932578,598244:13914581}). Among different types of neural networks, convolutional networks recently gained popularity for seismological applications because of their ability for automatic feature extraction and scaling to long input vectors, which are usually the case for earthquake signals; however, their sensitivity to un-normalized input data, makes their application for magnitude estimation challenging because amplitude information plays a key role (at least for scales like local magnitude).  To overcome this problem, we designed a network that mainly consisted of convolutional and recurrent layers where the convolutional layers do not have any activation function but are used only for dimensionality reduction and feature extraction (Figure~\ref{f-5df8cb49bcb0}).

\bgroup
\fixFloatSize{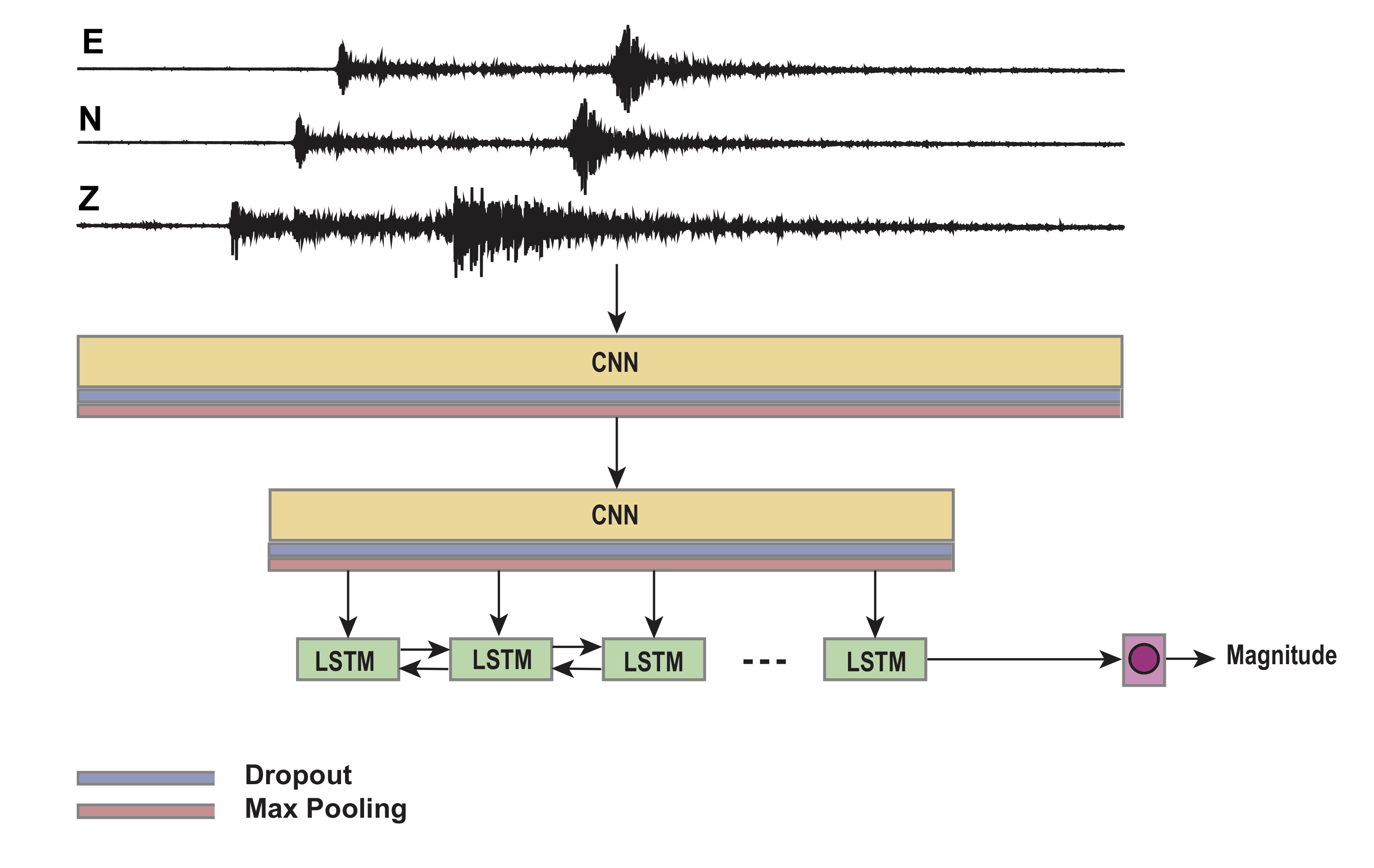}
\begin{figure*}[!htbp]
\centering \makeatletter\IfFileExists{images/f52c4650-5ec1-4c59-8644-0cd03a568843-ufig_1.jpg}{\includegraphics{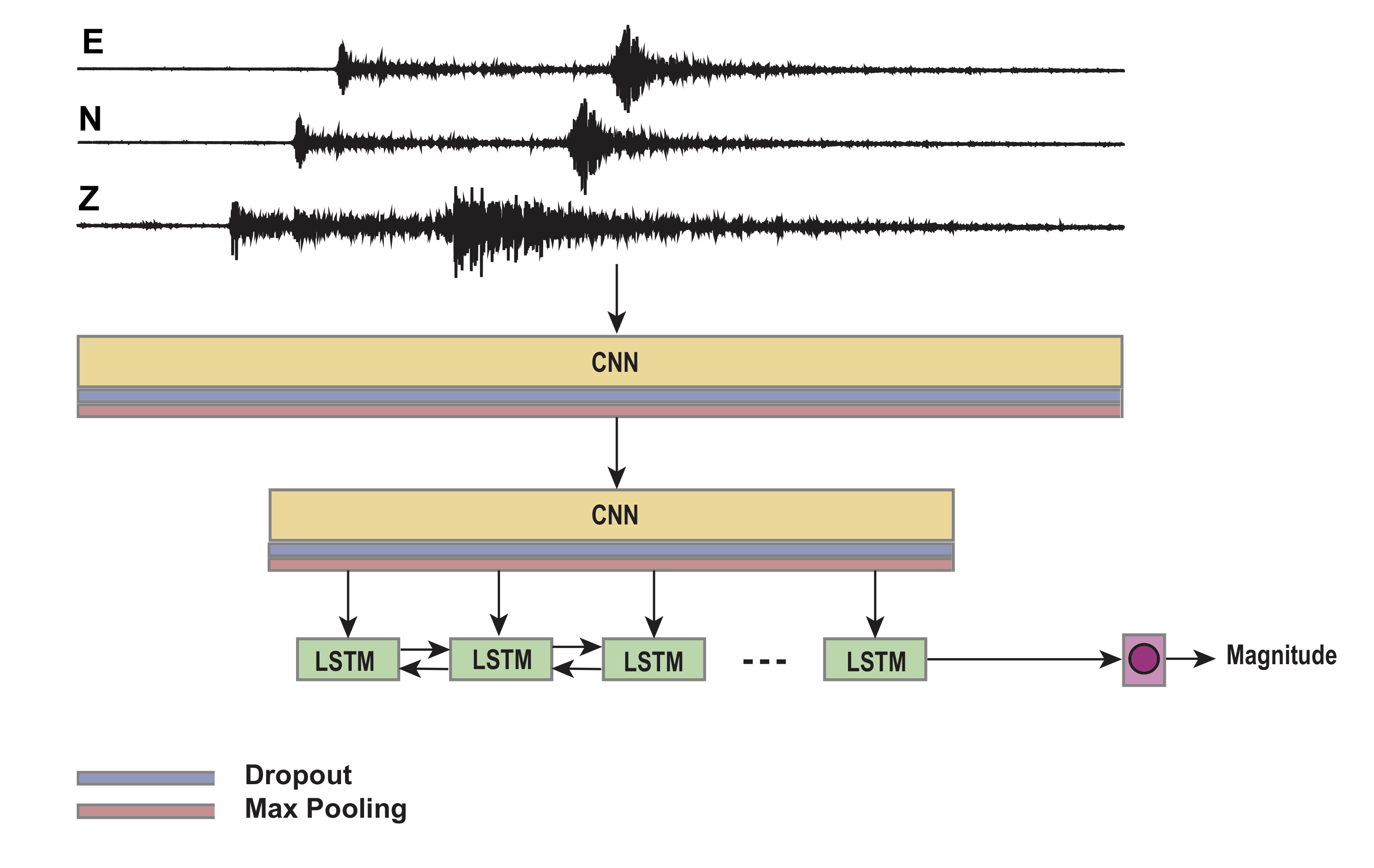}}{}
\makeatother 
\caption{{Network architecture. }}
\label{f-5df8cb49bcb0}
\end{figure*}
\egroup
The learning is mainly done in the Long-Short Term-Memory (LSTM) units and their following fully connected layers. Long-Short Term-Memory (LSTM) (\cite{598244:13914749}) are specific types of recurrent neural networks capable of retaining the temporal dependencies among the input elements during the training process. Hence, they are commonly used for modeling of sequential data similar to earthquake signals. A detailed description of LSTMs and their applications to earthquake data can be found in \unskip~\cite{598244:13914704}. The advantage of using LSTM units for magnitude estimation lies in their insensitivity to un-normalized inputs due to their gated mechanism. 

The input to the network are three-channel seismograms each 30 second (3000 smaples) long. Our network (Figure~\ref{f-5df8cb49bcb0}) consists of two convolutional layers (with 64 and 32 kernels of size 3 respectively) at its fore front, each followed by a dropout (\cite{598244:13915074}) and maxpooling (\cite{598244:13915075}) layer. Dropout layers are used for regularization. Each maxpooling layer reduces the dimension of the input data by a factor of 4 to facilitate the training speed at the bidirectional LSTM layer (with 100 units). At the end, the output of the LSTM layer is passed to a fully connected layer with one neuron and a linear activation to estimate the magnitude. Here, we trained the model with dropout rate of 0.2 by minimizing the mean square error. 
    
\section{Results}
We use a selected portion of STanford EArthquake Dataset (STEAD)\unskip~\cite{598244:14253409} for training of the network. STEAD is a global dataset of labeled seismograms including local earthquake and seismic noise waveforms.  Here, we only used {\texttildeapprox} 300,000 earthquake waveforms recorded at epicentral distances of less than 1 degree, for which their full waveforms (from 1 second before P until the end of the S coda) are equal or less than 30 seconds. The magnitude distribution of the events is shown in Figure~\ref{f-563eff501350}. All waveforms were band-passed filtered between 1.0-40.0 Hz and have signal-to-noise ration of greater than 20 db. To investigate the potential effects of various factors such as magnitude type, site effects, regional effects (using data in a specific geographical location), and site-dependent learning (using a limited number of stations) we divide the data into smaller subsets, each from 60K to 140K, and used 70\% of each subset for the training and 10\% and 20\% for the validation and testing respectively. We train the network using Adam optimizer (\cite{598244:13914703}) and automatically stop the training when validation loss doesn't decrease for 5 consecutive epochs to avoid overfitting.

\bgroup
\fixFloatSize{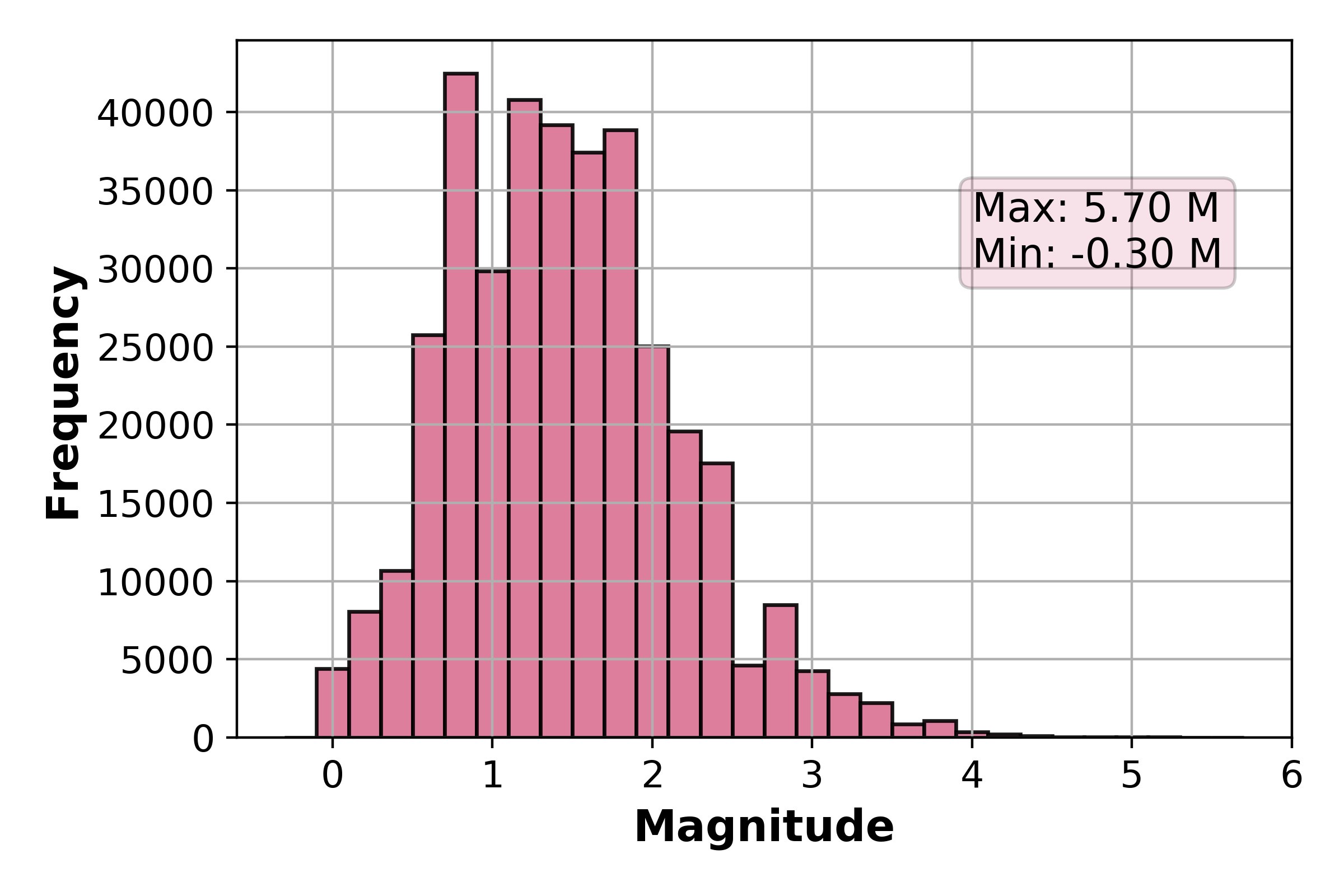}
\begin{figure*}[!htbp]
\centering \makeatletter\IfFileExists{images/5f8328d1-ed67-41bc-917b-e7cdca391989-ufig_2.jpg}{\includegraphics{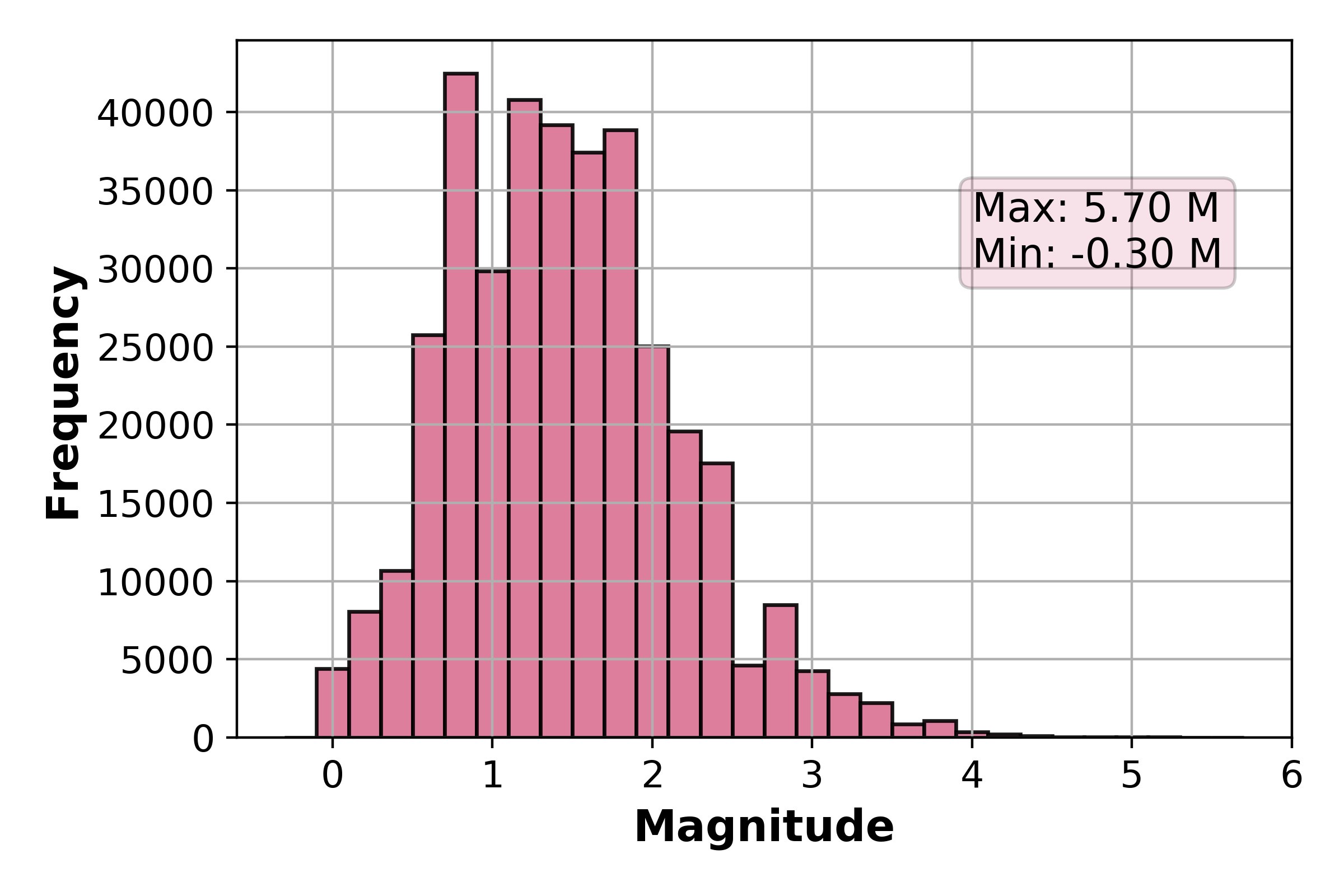}}{}
\makeatother 
\caption{{Magnitude distribution of events used for training and tests. }}
\label{f-563eff501350}
\end{figure*}
\egroup
Results are presented in Figure~\ref{f-0e400e2da790}. Overall the network is able to predict earthquake magnitudes with a mean error close to zero and standard deviation of {\texttildeapprox}0.2. The regression performance is worst at upper and lower bounds where fewer training/test samples are available for larger and smaller magnitudes size respectively. The network can predict both local and duration magnitudes with a reasonable accuracy which indicates its ability to learn both attenuation (interpreting the amplitudes with regards to the event-station distance) and duration, directly from the input waveform from a single station. We note that these estimates for local magnitude are based on global data recorded and reported by 98 monitoring networks around the world where different attenuation or calibration corrections might have been applied. To see if regional based models perform better, we build a model using only events in southern California. Although the coefficient of determination decreases slightly, due to relatively smaller size of the training set, we see only a small improvement in standard deviation of prediction error. 

In single-station estimates of the earthquake magnitudes, site amplification can play an important role. To check for a potential effect of this factor on our network's estimates, we build two separate models based on surface and borehole station data. Although performance deteriorates in the case of the surface stations, it is hard to associate this to the site amplification alone because surface data have higher noise levels which, as we will see later, has a direct impact on the magnitude estimates. 

We test whether a station-based model performs better by building a model using only stations (globally distributed) that have more than 1000 observations each. The best result is obtained by this model, suggesting a stronger effect of sites than regions on the learning performance.

\bgroup
\fixFloatSize{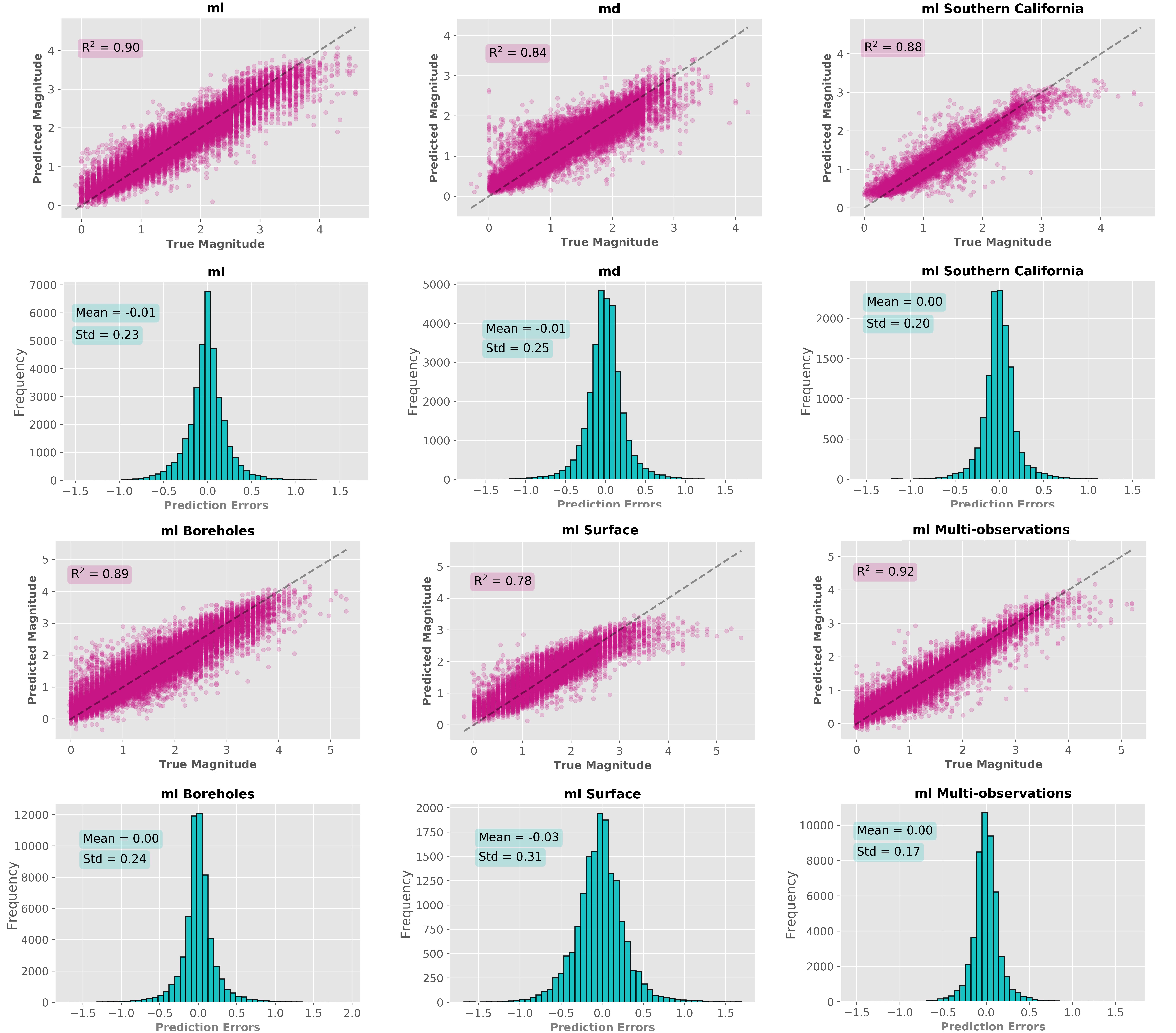}
\begin{figure*}[!htbp]
\centering \makeatletter\IfFileExists{images/ff28014e-c71e-407e-8661-817badcb7c58-ufig_3.jpg}{\includegraphics{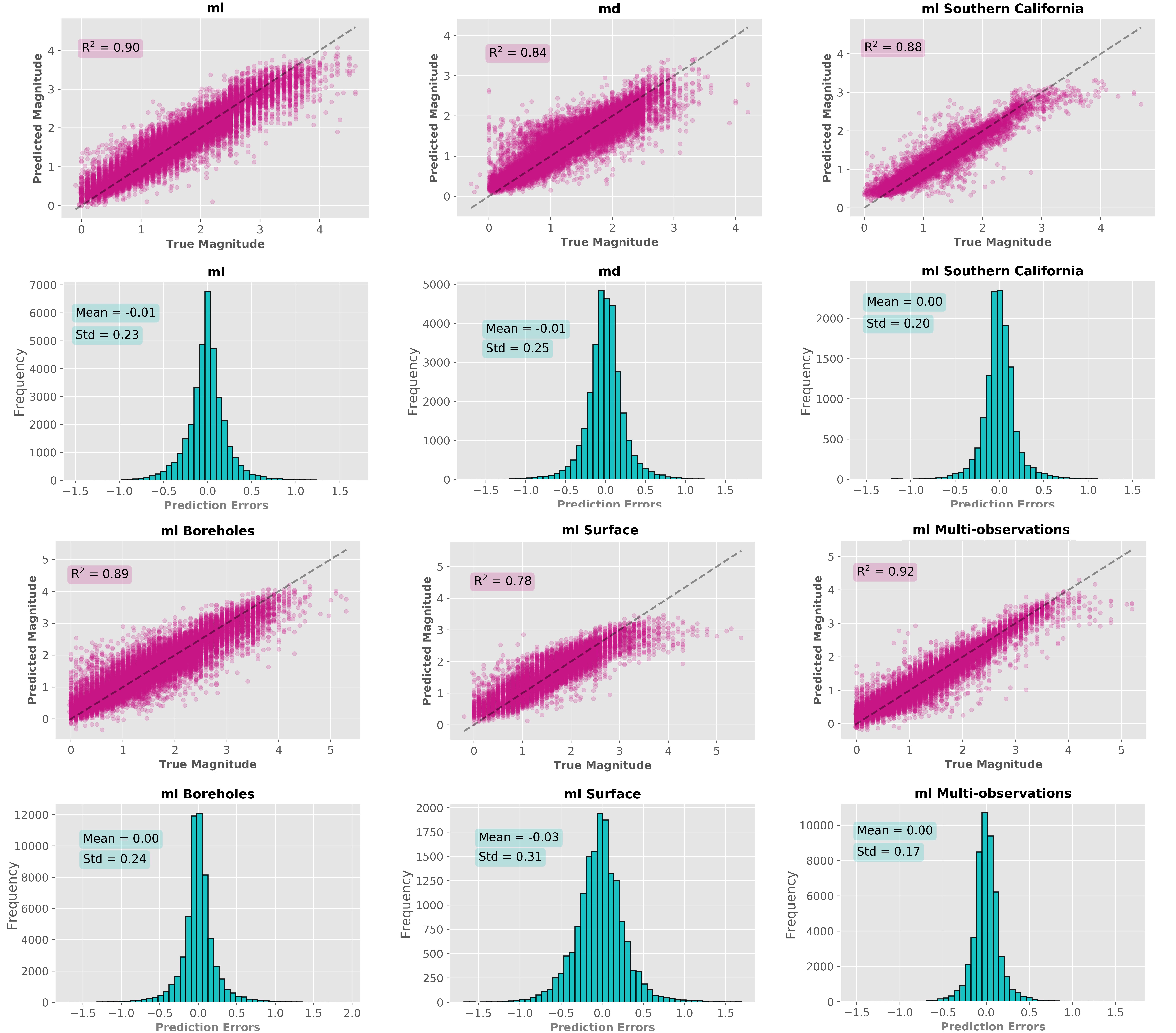}}{}
\makeatother 
\caption{{Prediction results on test sets for local magnitude (ml), duration magnitude (md), events occured in the Southern California, borehole stations, surface stations, and in a case where only stations with more than 1000 observations are used. }}
\label{f-0e400e2da790}
\end{figure*}
\egroup
The cataloged magnitudes used for training are basically averaged values over multiple stations for each event. To see how close our single-station estimates are to each other and the multi-station averaged estimations, we look at 311 events in our $M_L $ dataset where 4 or more observations (stations) are available for each event. Variation of single-station observations for each event are very small and in most cases within the ground truth range (Figure~\ref{f-4e745d354e52}). For each event, the predicted values for each station, averaged prediction, and ground truth are provided in the supplementary materials.

\bgroup
\fixFloatSize{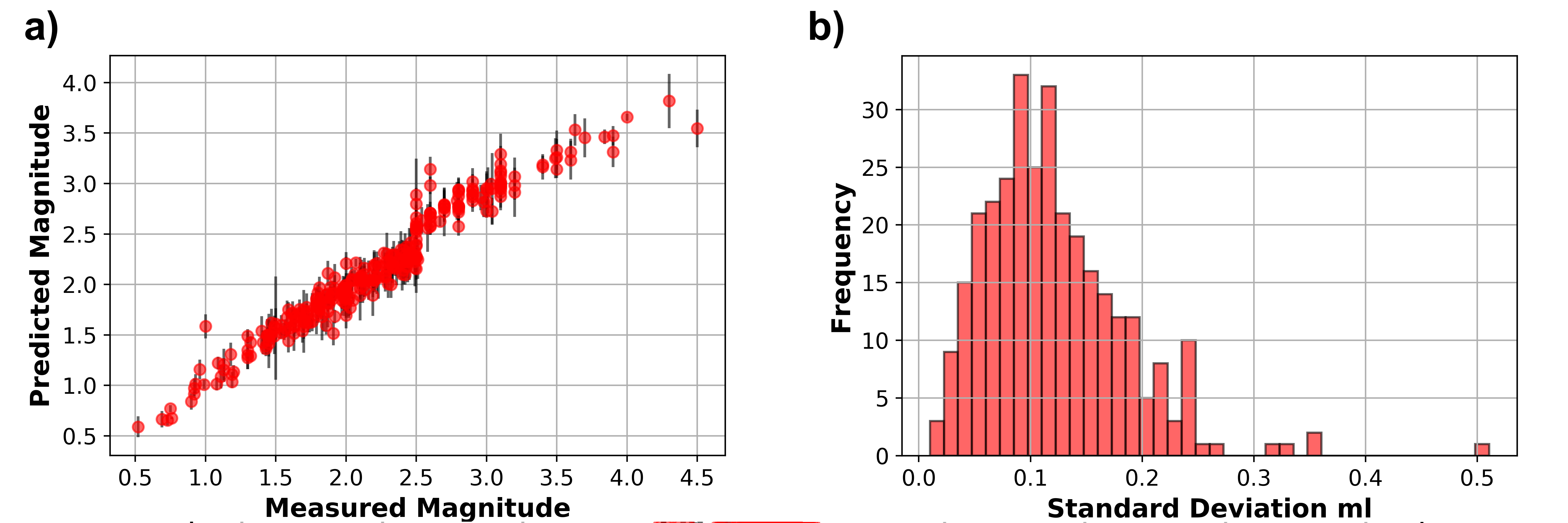}
\begin{figure*}[!htbp]
\centering \makeatletter\IfFileExists{images/eb37e347-84e5-469c-976f-db2957c4a78f-ufig_4.jpg}{\includegraphics{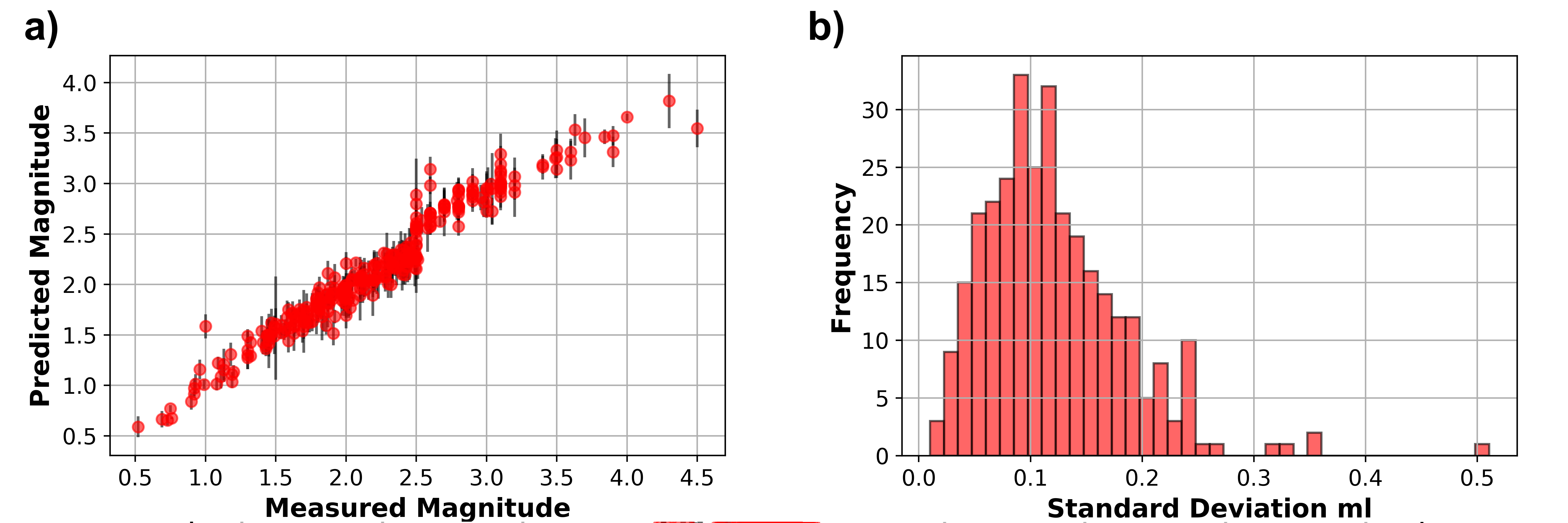}}{}
\makeatother 
\caption{{a) circles are averaged values over multiple station predictions where the error bar represents the standard deviation for single-station predictions of the event's magnitude. b) distribution of standard deviations of single-station estimations. }}
\label{f-4e745d354e52}
\end{figure*}
\egroup
We found that the signal-to-noise ratio has the grater impact on the performance of our regressor network (Figure~\ref{f-cde3474e1132}). Building a deeper network can be an effective solution to make the network less sensitive to the noise level; however, this requires more training data to prevent overfitting.

\bgroup
\fixFloatSize{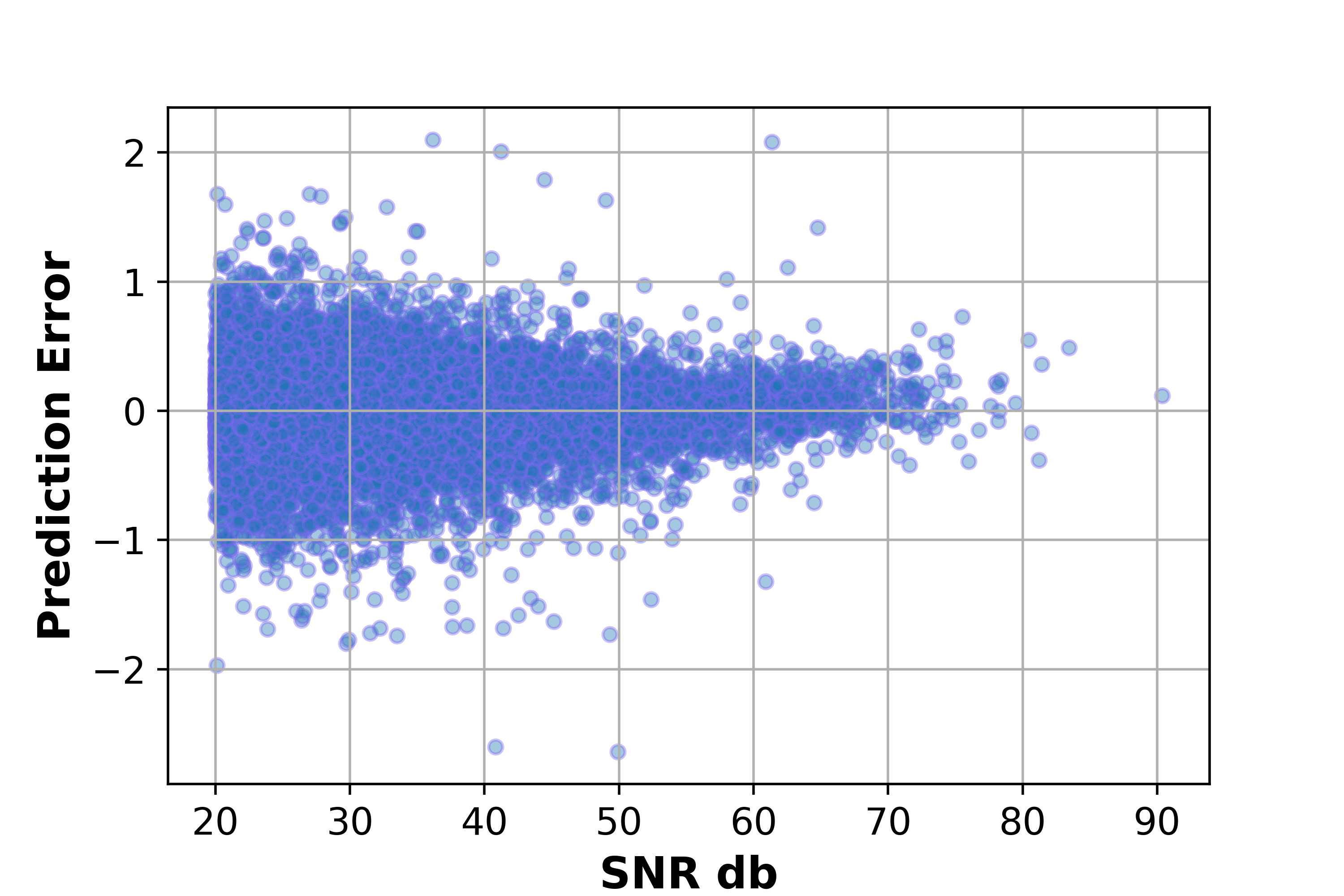}
\begin{figure*}[!htbp]
\centering \makeatletter\IfFileExists{images/16ff3153-b56c-41df-a99f-4376c1725add-ufig_5.jpg}{\includegraphics{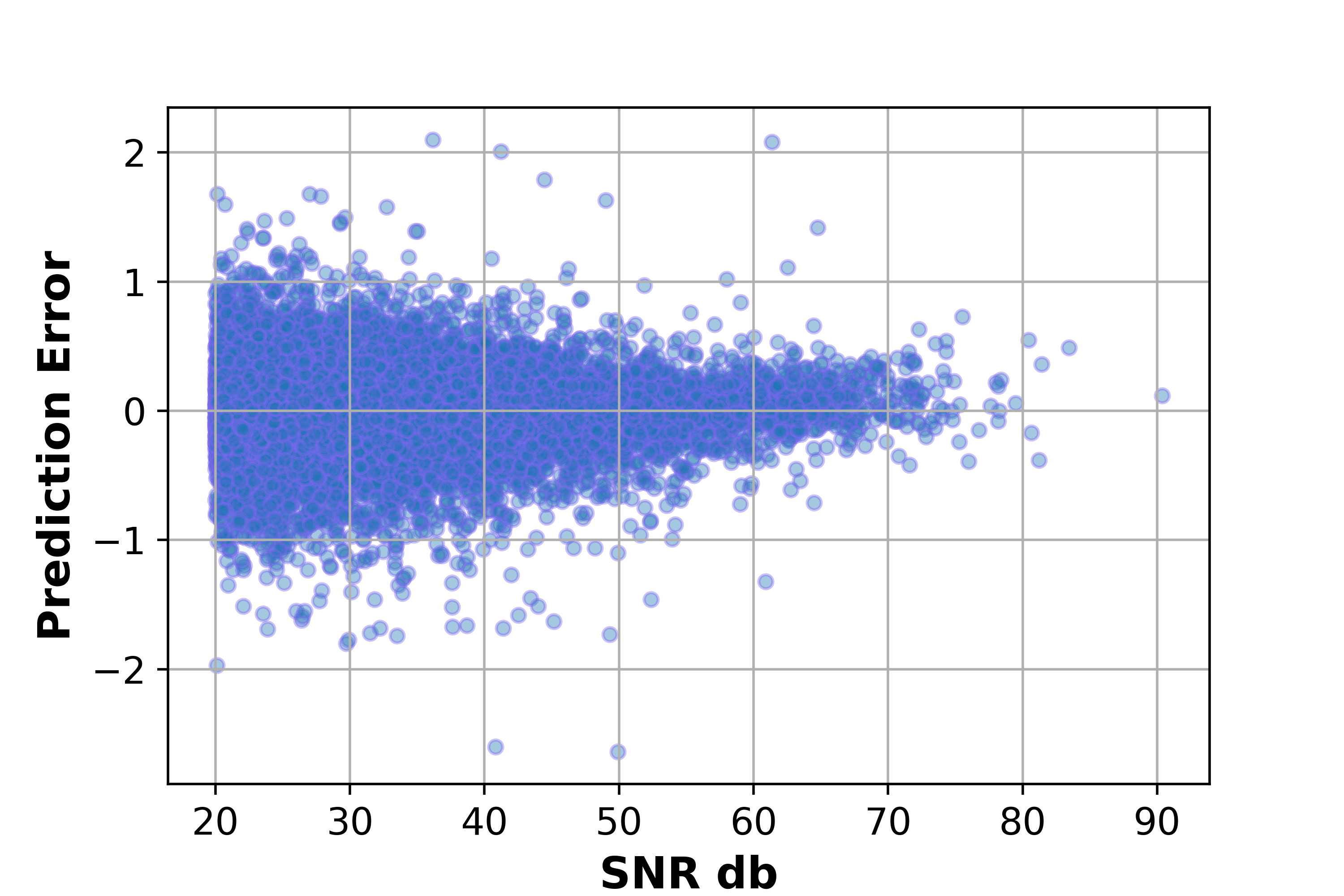}}{}
\makeatother 
\caption{{Prediction errors as a function of signal-to-noise ratio. For local magnitude estimates. }}
\label{f-cde3474e1132}
\end{figure*}
\egroup

\section{Conclusions}
 We showed that neural networks can learn general relations for estimating earthquake magnitudes directly from raw single-station waveforms. These estimates can be based on different magnitude types provided enough training data are available. Our results suggest site-specific learning can be an effective strategy for improving the performance, while region-based training might not be that important. Although we obtained better results using borehole stations compared with the surface stations, it is hard to conclude that site amplification effects alone are responsible for these results because a strong effect of noise level is also observable. The proposed method can provide a fast estimation of earthquake magnitude from raw seismograms observed at single stations. This has a variety of potential applications from routine earthquake monitoring to early warning systems.

\bibliographystyle{naturemag}

\bibliography{\jobname}\section*{Acknowledgements}SMM was partially support by Stanford Center for Induced and Triggered Seismicity during this project. GCB was supported by AFRL under contract number FA9453-19-C-0073. Seismic waveform data are available from Incorporated Research Institutions for Seismology (IRIS) at \BreakURLText{http://ds.iris.edu/ds/.} 

\newpage 

\end{document}